%% file: main.tex
\documentclass[sigconf,9pt]{acmart}
\AtBeginDocument{%
  \providecommand\BibTeX{{%
    \normalfont B\kern-0.5em{\scshape i\kern-0.25em b}\kern-0.8em\TeX}}}

\setcopyright{acmcopyright}

\copyrightyear{2023}
\acmYear{2023}
\setcopyright{acmlicensed}\acmConference[GPGPU '23]{15th Workshop on General Purpose Processing Using GPU}{February 25, 2023}{Montreal, Canada}
\acmBooktitle{15th Workshop on General Purpose Processing Using GPU (GPGPU '23), February 25, 2023, Montreal, Canada}
\acmPrice{15.00}
\acmDOI{10.1145/3589236.3589242}
\acmISBN{979-8-4007-0776-6/23/02}

\input{myliststyle}
\begin{document}

\title{Exploiting Scratchpad Memory for Deep Temporal Blocking}
\subtitle{A case study for 2D Jacobian 5-point iterative stencil kernel (j2d5pt)}

\settopmatter{authorsperrow=4}
\author{Lingqi Zhang}
\affiliation{%
  \institution{Tokyo Tech}
  \institution{AIST} 
  \country{Japan}
}
\author{Mohamed Wahib}
\affiliation{%
  \institution{RIKEN R-CCS}
  \country{Japan}
}
\author{Peng Chen}
\affiliation{%
  \institution{AIST}
  \institution{RIKEN R-CCS}
  \country{Japan}
}
\author{Jintao Meng}
\affiliation{%
  \institution{SIAT}
  \country{China}
}
\author{Xiao Wang}
\affiliation{%
  \institution{ORNL}
  \country{USA}
}
\author{Toshio Endo}
\affiliation{%
  \institution{Tokyo Tech}
  \country{Japan}
}
\author{Satoshi Matsuoka}
\affiliation{%
  \institution{RIKEN R-CCS}
  \institution{Tokyo Tech}
  \country{Japan}
}

\renewcommand{\shortauthors}{Lingqi Z. et al.}
\begin{abstract}

General Purpose Graphics Processing Units (GPGPU) are used in most of the top systems in HPC. The total capacity of scratchpad memory has increased by more than 40 times in the last decade. However, existing optimizations for stencil computations using temporal blocking have not aggressively exploited the large capacity of scratchpad memory. This work uses the 2D Jacobian 5-point iterative stencil as a case study to investigate the use of large scratchpad memory. Unlike existing research that tiles the domain in a thread block fashion, we tile the domain so that each tile is large enough to utilize all available scratchpad memory on the GPU. Consequently, we process several time steps inside a single tile before offloading the result back to global memory. Our evaluation shows that our performance is comparable to state-of-the-art implementations, yet our implementation is much simpler and does not require auto-generation of code.

 
\end{abstract}



\keywords{GPGPU, Temporal Blocking, Iterative Stencil Solvers}

\ccsdesc[500]{Computing methodologies}
\ccsdesc[500]{Computing methodologies~Vector / streaming algorithms}
\ccsdesc[500]{Computing methodologies~Massively parallel algorithms}
\maketitle



\section{Introduction}
When observing the previous generations of GPUs, Nivida GPUs for instance, there is a clear trend of increase in the cache capacity. Especially the volume of scratchpad memory (or shared memory in CUDA~\cite{nvidia2019programming}) increased from $720$ KB in K20 (2013) to $17.30$ MB in A100 (2020). The latest H100 (2023) GPU even pushes max usable shared memory to be $29.83$ MB to more than $200$ KB per stream multiprocessor(SM).

GPU optimizations that are commonly used in HPC applications were designed mostly assuming that scratchpad memory is not larger than $100$ KB per stream multiprocessor~\cite{rawat2018optimization}. There is a potential in leveraging the untapped scratchpad memory to aggressively optimize for data locality. 

In this work, we use a case study kernel commonly used in HPC applications, namely 2D Jacobian 5-point iterative stencil, to fully take advantage of the scratchpad memory for tiling data in an unusual way. More specifically, we run each of the tiles in a serial fashion one after the other while aggressively using the shared memory to run each tile entirely from shared memory. We use device-wide synchronization to resolve the spatial dependency between thread blocks. 
We demonstrate a new approach to leverage the large capacity of shared memory by proposing a temporal blocking stencil scheme that optimizes for peak data locality, i.e. running the entire problem from shared memory. Our method is much simpler than complex temporal blocking schemes; iterative kernels that use our methods can be manually written, unlike complex temporal schemes that require auto-generation of code.


\section{Related work}
Temporal blocking~\cite{DBLP:conf/cgo/MatsumuraZWEM20,rawat2018domain} tiles the domain and processes the domain with in combined time steps. Due to space limitations, we mainly review StencilGen~\cite{rawat2018domain} and AN5D~\cite{DBLP:conf/cgo/MatsumuraZWEM20}. Both works used 2.5D or 3.5D tiling and relied on code auto generation for performance optimization. In addition, they relied on overlapped tiling within thread blocks. They did not exploit the inter thread block data exchange pattern. Regarding the usage of scratchpad memory, StencilGen stores all combined time steps in scratchpad memory; AN5D uses scratchpad memory conservatively for double buffer. As a result, in the j2d5pt double-precision kernel. StencilGen and AN5D consumed about $4.32$ MB and $0.864$ MB scratchpad memory, respectively. So, both AN5D and StencilGen left most of the scratchpad memory untapped, and are overly complex to implement.

\section{Deep temporal blocking (DTB)}
\lstset{
 	language = C++, breaklines = true, breakindent = 10pt, lineskip={-1pt}, basicstyle = \rmfamily\scriptsize, commentstyle = {\itshape \color[cmyk]{1,0.4,1,0}}, classoffset = 0, keywordstyle = {\bfseries \color[cmyk]{0,1,0,0}}, stringstyle = {\ttfamily \color[rgb]{0,0,1}}, frame = trbl, framesep=0pt, numbers = left, stepnumber = 1, xrightmargin=12pt, xleftmargin=0pt, numberstyle = \tiny, tabsize = 1, captionpos = t, directivestyle={\color{black}},  emph={int,char,double,float,unsigned, int3, float4, float2}, emphstyle={\color{blue}},
}
\lstset{escapeinside={<@}{@>}}

\subsection{Basic function}
\begin{figure}[t]
\centering
\begin{minipage}[c]{0.5\textwidth}
\begin{lstlisting}[caption = {Pseudo code for j2d5pt stencil kernel function}, label=Fig:codeperk2d5pt]
//kernel function don't assume whether ptr_in and ptr_out is in device memory or scratchpad memory
__device__ void j2d5pt(ptr_in, ptr_out, loc_x, loc_y){ 
    x = threadIdx.x; 
    t[ILP+2]; //ILP: instruction level parallel
    for(y=0; y< ILP+2; y++){
        t[y]=ptr_in[x, y+ind_y-1]; 
        for( y=0; y< ILP; y++){
            result[y]=ptr_in[x+loc_x-1,y+1+loc_y]*W
            +ptr_in[x+loc_x+1,y+1+loc_y]*E 
            +t[y-1+1]*S 
            +t[y+1]*  C 
            +t[y+1+1]*N;
        }
    }
    for( y=0; y< ILP; y++){
       ptr_out[x+loc_x,y+loc_y] = result[y];
    }
}
\end{lstlisting}
\end{minipage}
\end{figure}
Listing~\ref{Fig:codeperk2d5pt} shows the base kernel function we used in this case study. We only modified the input and output pointer location to use scratchpad memory. In this kernel, we move the time loop from the host side to he be inside the kernel. Next, we tile the domain of the problem spatially and run the tiles in a serial fashion. For each tile, we run it entirely to completion, over all its time steps, before we start on the next tile.
\subsection{Dependency Between Thread Blocks}
We use the CUDA grid-level barrier to ensure that each thread block can exchange the halo region correctly. 
We use the bulk synchronous parallel (BSP) model.
\subsection{Processing the Tiles in Order}
\begin{figure}[t]
\centering
\includegraphics[width=\linewidth]{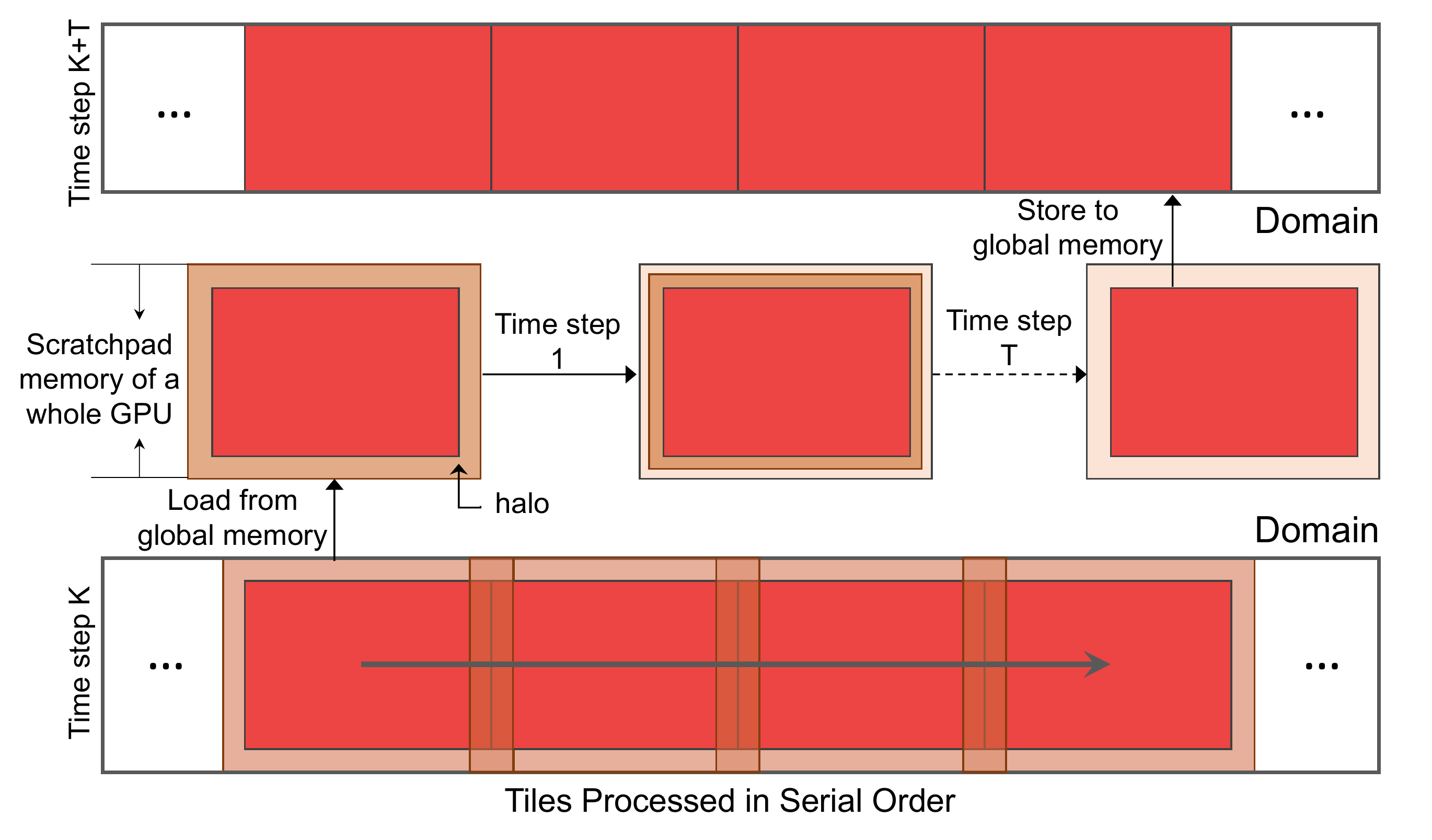}
\caption{\label{fig:exestencil} How DTB processes the tiles. DTB loads the tile to populate the scratchpad memory with the input, processes T time steps, and then stores the results to the output address. DTB processes tiles in a serial order.
}
\end{figure}

After we load a tile into the scratchpad memory, we process the tile for several time steps (temporal blocking) before moving to the next tiling. Figure~\ref{fig:exestencil} shows the process. 



\section{Evaluation}

\begin{figure}[t]
\centering
 \includegraphics[width=0.9\linewidth]{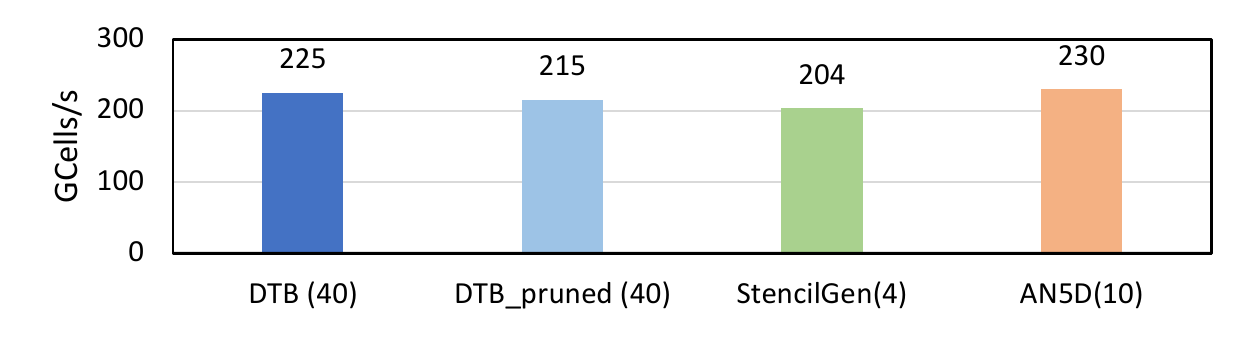}
\caption{\label{fig:cmp} Comparing the performance of DTB with other state-of-the-art temporal blocking implementations (SOTAs), i.e., StencilGen~\cite{rawat2018domain} and AN5D~\cite{DBLP:conf/cgo/MatsumuraZWEM20}. The temporal blocking depth (number of time steps) is marked inside the parentheses. 
DTB runs a $8592\times8328$ domain. DBT\_pruned, StencilGen, and AN5D run $8192^2$ domain size. We use the valid domain to evaluate the performance. 
DTB shows comparable performance with other SOTAs. 
}
\end{figure}

We compare DTB with StencilGen~\cite{rawat2018domain} and AN5D~\cite{DBLP:conf/cgo/MatsumuraZWEM20}, 
the state-of-the-art implementations for temporal blocking for stencils (a domain size of $8192^2$). We used $8592\times8328$ to run the DTB. We also report a pruned version that considers $8192^2$ as a valid domain size. 
Figure~\ref{fig:cmp} shows the result: the performance of DTB is comparable to that of state-of-the-art temporal blocking implementations (SOTAs).

\section{Conclusion}
In this work, we discuss a case study on the use of scratchpad memory for DTB on the j2d5pt stencil. Instead of applying a complex temporal blocking implementation, we just tile the domain so that each tile fully occupies the scratchpad memory. Evaluation shows that DTB is compatible with other SOTAs. 
 We anticipate that DTB could perform even better on a larger scratchpad memory architecture, which would be explored in future work.




\begin{acks}
This work was supported by JSPS KAKENHI under Grant Number JP21K17750.
This paper is based on results obtained from a project, JPNP20006, commissioned by the New Energy and Industrial Technology Development Organization (NEDO).
This research used resources at the Oak Ridge Leadership Computing Facility, a DOE Office of Science User Facility operated by the Oak Ridge National Laboratory. The authors wish to express their sincere appreciation to Jens Domke, Aleksandr Drozd, Emil Vatai and other RIKEN R-CCS colleagues for their invaluable advice and guidance throughout the course of this research. Finally, the first author would also like to express his gratitude to RIKEN R-CCS for offering the opportunity to undertake this research in an intern program. 
\end{acks}



\bibliographystyle{ACM-Reference-Format}
\bibliography{lingqi}

\end{document}

%% file: myliststyle.tex
\usepackage{listings}
\usepackage{color}
\definecolor{codegreen}{rgb}{0,0.6,0}
\definecolor{codegray}{rgb}{0.5,0.5,0.5}
\definecolor{codepurple}{rgb}{0.58,0,0.82}
\definecolor{backcolour}{rgb}{0.95,0.95,0.92}
 
\usepackage{caption}

\lstdefinestyle{mystyle}{
    backgroundcolor=\color{white},   
    commentstyle=\color{codegreen},
    keywordstyle=\color{magenta},
    numberstyle=\tiny\color{codegray},
    stringstyle=\color{codepurple},
    basicstyle=\scriptsize,
    breakatwhitespace=false,         
    breaklines=true,                 
    captionpos=b,                    
    keepspaces=true,                 
    numbers=left,                    
    numbersep=5pt,                  
    showspaces=false,                
    showstringspaces=false,
    showtabs=false, 
    tabsize=2
}
 